\documentstyle[aps,prb,epsf,floats]{revtex}
\begin{document}
\twocolumn[\hsize\textwidth\columnwidth\hsize\csname@twocolumnfalse%
\endcsname
\title{Spin and charge order in Mott insulators \\ and $d$-wave superconductors}
\author{Subir Sachdev}
\address{Department of Physics, Yale University, P.O. Box 208120, New
Haven, CT 06520-8120,\\
and Department of Physics, Harvard University, Cambridge MA 02138}
\date{August 15, 2001}

\maketitle

\begin{abstract}
We argue that aspects of the anomalous, low temperature, spin and
charge dynamics of the high temperature superconductors can be
understood by studying the corresponding physics of undoped Mott
insulators. Such insulators display a quantum transition from a
magnetically ordered N\'{e}el state to a confining paramagnet with
a spin gap; the latter state has bond-centered charge order, a low
energy $S=1$ spin exciton, confinement of $S=1/2$ spinons, and a
free $S=1/2$ moment near non-magnetic impurities. We discuss how
these characteristics, and the quantum phase transitions, evolve
upon doping the insulator into a $d$-wave superconductor. This
theoretical framework was used to make a number of predictions for
STM measurements and for the phase diagram of the doped Mott
insulator in an applied magnetic field.
\begin{center}{\tt Talk at SNS 2001.\\
Conference on the Spectroscopies of Novel Superconductors,\\
Chicago, May 13-17, 2001.}\end{center}
\end{abstract}
\pacs{PACS numbers:}  ]


\section{Introduction}
\label{intro}

Although there is general agreement that the ground state of the
high temperature superconductors is a conventional BCS $d$-wave
superconductor, a number of low temperature properties do not fit
easily into this framework, especially in the underdoped regime.
Among these are ({\em a\/}) the presence of a $S=1/2$ moment near
non-magnetic Zn or Li impurties in the underdoped superconductor
\cite{bobroff,fink,alloulold}, ({\em b\/}) low energy collective
spin excitations (a $S=1$ spin exciton) at incommensurate
wavevectors near $(\pi,\pi)$, and ({\em c\/}) instabilities to
various co-existing spin and charge density wave states
(``stripes'') in certain materials. While it is certainly possible
to devise microscopic models, and a corresponding Hartree-Fock
treatment, to generate any of these physical properties in the
superconducting state, such a procedure is somewhat ad hoc and has
limited predictive power.

\begin{figure}
\epsfxsize=3.5in \centerline{\epsffile{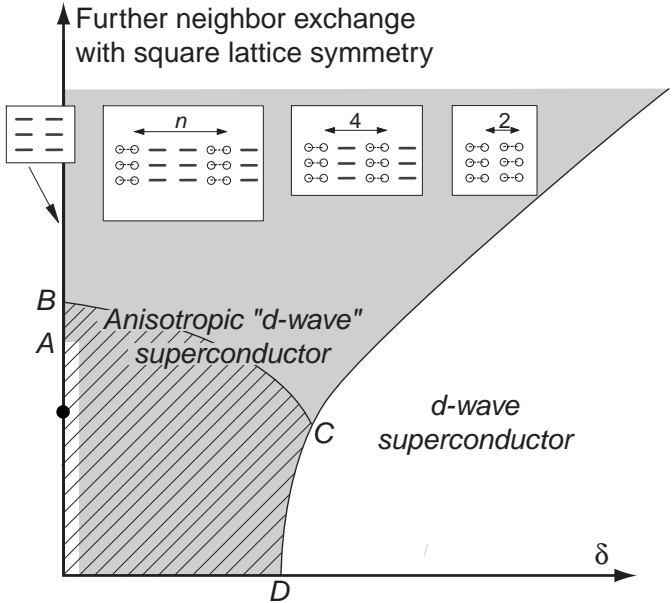}}
\caption{Schematic $T=0$ phase diagram (adapted from
Ref.~\protect\onlinecite{vojtaprl}) for the high temperature
superconductors as a function of a ratio of the near neighbor
exchange interactions and the hole concentration, $\delta$; {\em
e.g.\/} the vertical axis could be $J_2/J_1$, the ratio of the
first to second neighbor exchange. The shaded region has charge
order {\em i.e.} at least one of the order parameters,
$\phi_{x,y}$, in (\protect\ref{rho}) is non-zero. The hatched
region has broken spin-rotation symmetry with $\langle \vec{S}
\rangle \neq 0$, and at least one of the order parameters
$\Phi_{x,y\alpha}$ in (\protect\ref{spin}) is non-zero; $\langle
\vec{S} \rangle =0$ elsewhere. The unfrustrated, insulating
antiferromagnet with long-range N\'{e}el order is indicated by the
filled circle. At $\delta=0$, there is an onset of charge order
above the point A, while spin-rotation invariance is restored
above B (for a review of the $\delta=0$ physics, see
Ref.~\protect\onlinecite{qaf}). The magnetic ordering quantum
transition for $\delta>0$ along the line BC is expected to be in
the same universality class as that at the point B at $\delta=0$,
and is discussed as case (I) in Section~\ref{qc2}. The magnetic
ordering transition along the line CD differs slightly and is
discussed as case (II) in Section~\ref{qc2}. Neither of these
theories apply to the multicritical point C, in the vicinity of
which the transition is expected to be first
order~\protect\cite{zachar}. It is not know whether doping the
unfrustrated antiferromagnet (denoted by the filled circle) leads
to a system which crosses line BC or CD. The nature of the charge
orders as determined by the computations of
Ref.~\protect\onlinecite{sr,vojtaprl,lt22,kwon} are indicated at
the top of the figure; numerous transitions, within the gray
shaded region, in the nature of the charge ordering are not shown.
The ground state at very low non-zero doping is an insulating
Wigner crystal and there is subsequently a
insulator-to-superconductor transition; superconductivity is
present over the bulk of the $\delta >0$ region. The central idea
behind our approach is that many essential aspects the spin
excitation spectrum of the insulating, paramagnetic region
($\delta=0$, $\langle \vec{S}\rangle =0 $) lead to a simple and
natural description of the analogous properties of the $d$-wave
superconductor.} \label{fig1}
\end{figure}

\begin{figure}
\epsfxsize=3.5in \centerline{\epsffile{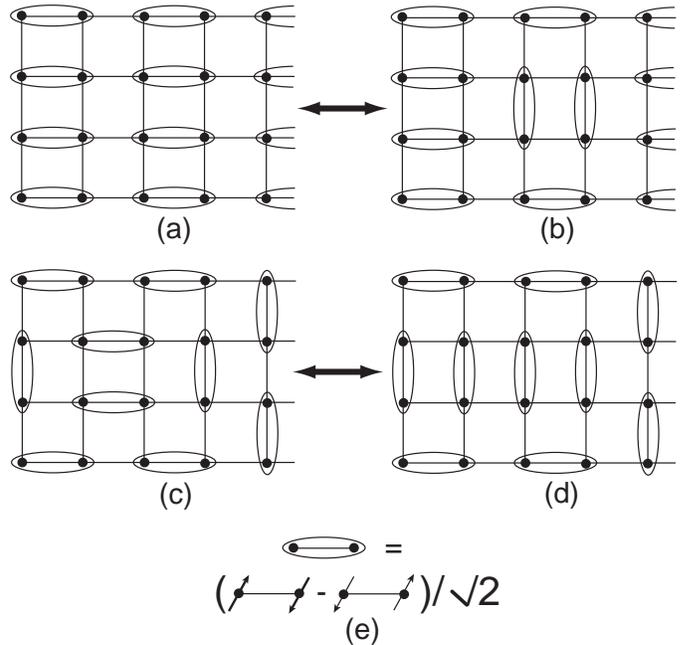}} \caption{
Representation of antiferromagnetic spin correlations in the
paramagnetic Mott insulator in the language of the quantum dimer
model\cite{rk}. Each picture is a snapshots of nearest-neighbor
singlet valence bonds, represented by ellipses as indicated in
(e). The states (a) and (b) (and similarly, the states (c) and
(d)) are connected to each other by resonance around a plaquette.
The state in (a) has the maximal number of plaquettes around
which such resonance can occur, and this lowers its energy. It
has been argued~\cite{qaf} that this effect leads to a broken
lattice symmetry in two dimensions, and the resulting ground
state has bond-centered charge order. A plausible candidate for
such a state is the one with spin-Peierls order---this state
breaks the square lattice symmetry in the same pattern as the
state in (a). The presence of this broken symmetry will also
modulate the bond charge density ({\em i.e.} the charge density
on an orbital which is a linear combination of the orbitals on
the two sites at the ends of the link) with period 2.}
\label{fig2}
\end{figure}

\begin{figure}
\epsfxsize=2.5in \centerline{\epsffile{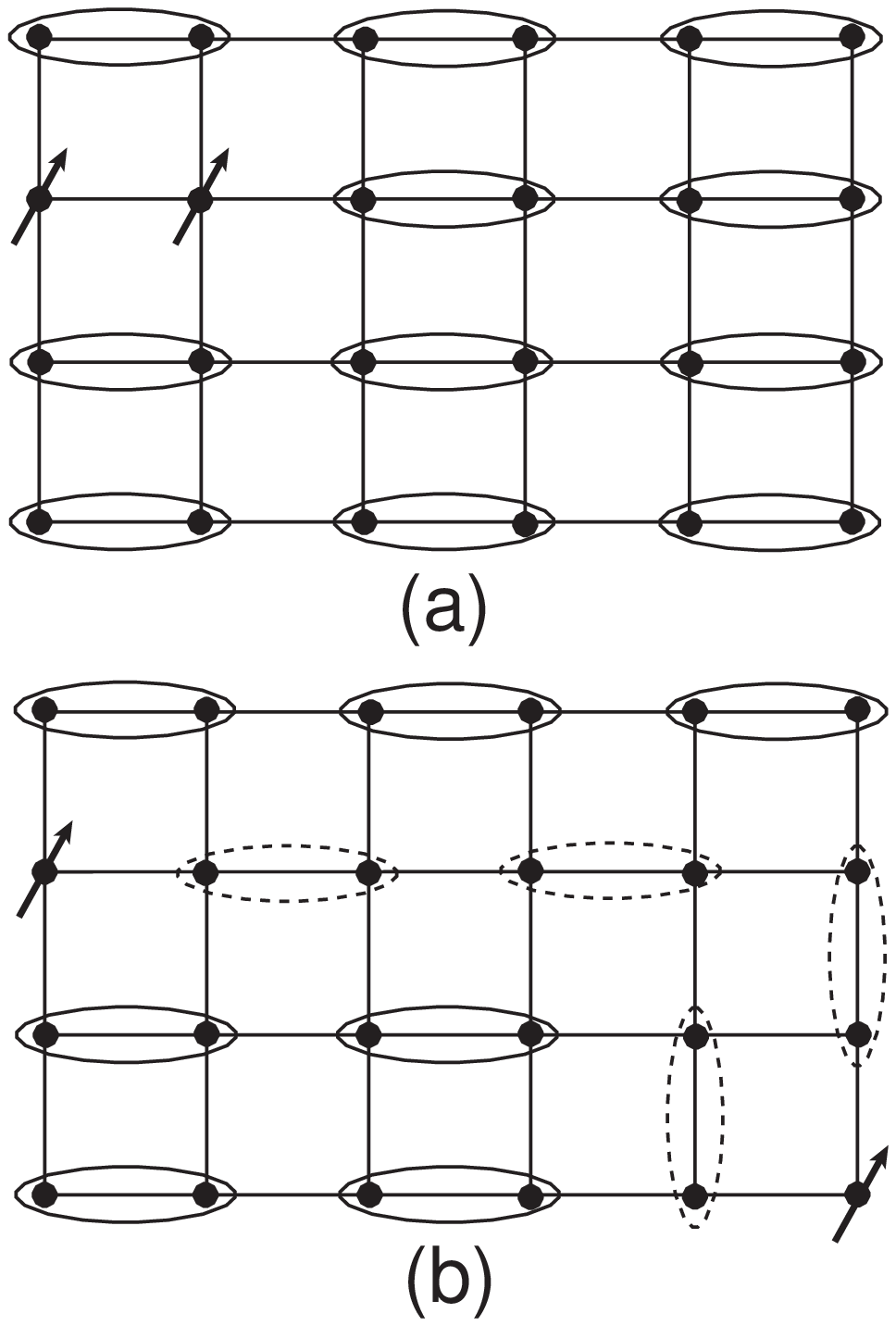}} \caption{(a)
Cartoon picture of the bosonic $S=1$ exciton of the insulating
paramagnet--a similar picture is also expected to apply in the
superconductor. (b) Fission of the $S=1$ excitation into two
$S=1/2$ spinons. Notice that the spinons are connected by  a
``string'' of valence bonds (denoted by dashed lines) which are
not able to resonate with their environment: this string costs a
certain energy per unit length and leads to the confinement of
spinons.} \label{fig3}
\end{figure}

\begin{figure}
\epsfxsize=2.5in \centerline{\epsffile{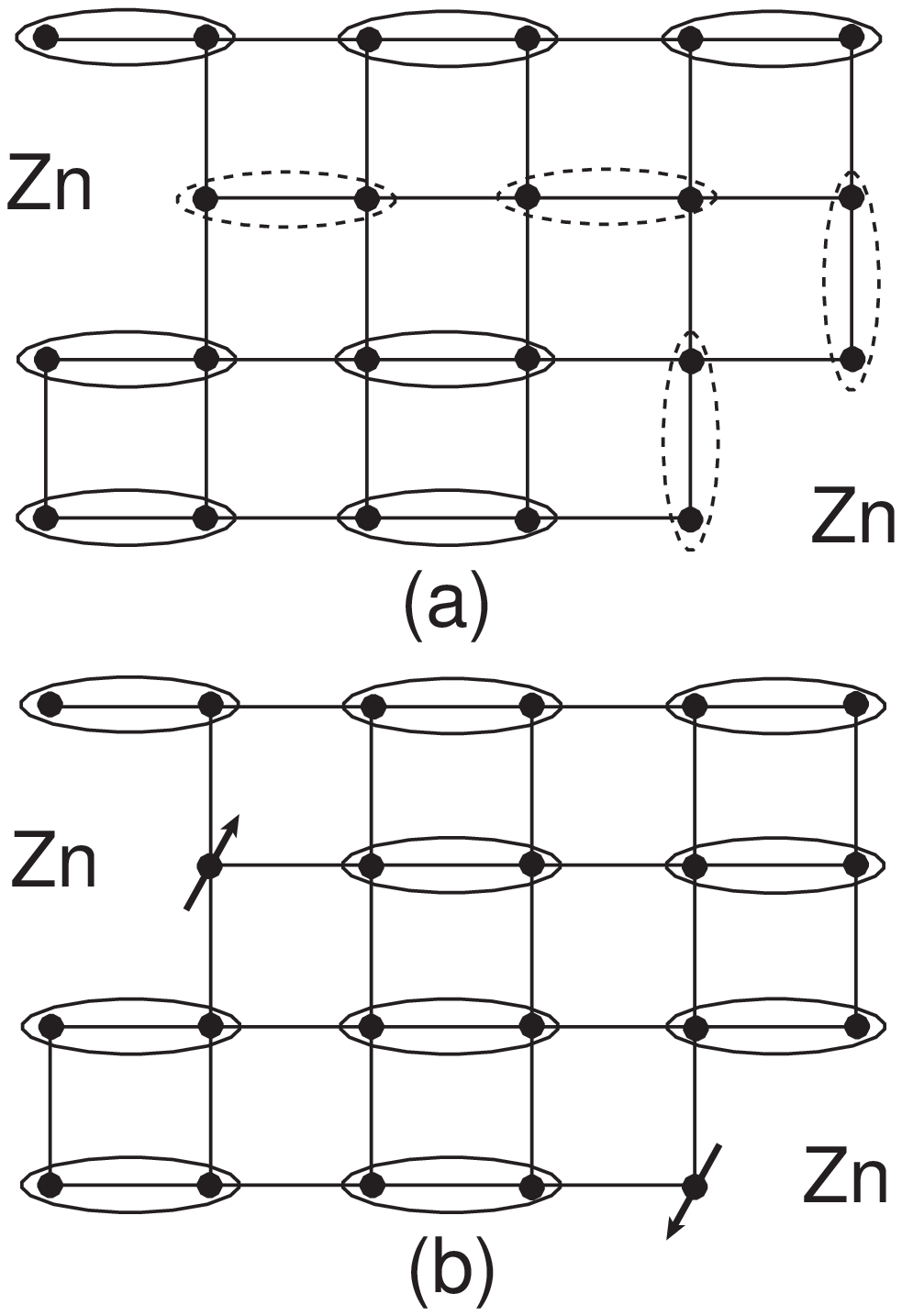}}
\caption{Confinement of a $S=1/2$ moment near a non-magnetic Zn
impurity~\protect\cite{icmp}. We consider two Zn impurities and
imagine moving them infinitely far apart. In the trial state in
(a), all Cu spins are paired into singlet bonds; while this is
preferable when the Zn impurities are not too far, there is a
string of valence bonds connecting them which are not able to
resonate, and this will eventually lead to the preference of the
structure in (b). In this case there is no string defect in the
spin-Peierls order, but each impurity has a ``free'' moment in its
vicinity. There is of course a very weak residual coupling
between these moments even when they are far apart, and so the
lowest energy states will be a singlet and triplet separated by a
gap which decreases exponentially with the separation between the
impurities. This should be contrasted with the case when moments
do not form around each impurity (as may be expected in a
deconfined paramagnet) when the gap above the ground state
saturates at a fixed finite value as the impurities move far
apart.} \label{fig4}
\end{figure}

Following an early suggestion of Anderson~\cite{pwa}, we explore
here the idea that these anomalous properties are related to
fundamental characteristics of the undoped Mott insulator, and
argue that this leads to a deeper understanding of the
physics~\cite{science}. The strategy can be summarized by
referring to the phase diagram in Fig~\ref{fig1}.
The first step in our theoretical program is an understanding of
paramagnetic Mott insulators which are `near' the N\'{e}el state
{\em i.e.} those obtained across a continuous quantum transition
at which N\'{e}el order is destroyed (this corresponds to moving
vertically along the $\delta=0$ line in Fig~\ref{fig1}). There has
been much work on this question in the last decade, and we refer
the reader to a recent paper by the author~\cite{qaf} for a
discussion of recent progress along with a review of earlier work.
The main result we focus on here is the prediction~\cite{rs1,sr}
that the appropriate paramagnetic Mott insulator has bond-centered
charge order and confinement of spinons. Fig~\ref{fig2} contains a
sketch of a likely bond-ordered state, along with a simple
physical argument for its selection.
A correspondingly simple argument for the confinement of $S=1/2$
spinons in such a state is sketched in Fig~\ref{fig3}.
The elementary excitation appears naturally as a $S=1$ exciton
indicated in Fig~\ref{fig3}. (This exciton will appear as a
`resonance peak' in the neutron scattering cross-section, and the
first prediction that such a peak would appear after the loss of
magnetic order in {\em doped\/} antiferromagnets (with or without
spin-Peierls order) was made in Ref.~\onlinecite{CSY}.) Another
crucial property of the confining paramagnetic insulator is the
prediction~\cite{rs1,fink} of a free $S=1/2$ moment near each
non-magnetic impurity: a simple argument for this appears in
Fig~\ref{fig4}.
The reader should already notice a strong similarity between the
properties of the paramagnetic Mott insulating state we have just
discussed and the properties ({\em a\/})-({\em c\/}) of the
underdoped superconductor. We will amplify on this connection in
the following sections, which contain a review of recent work,
accompanied by a discussion of some relevant experiments.

\section{Non-magnetic impurities}

We have already mentioned the recent observation of $S=1/2$
moments near non-magnetic impurities at low $T$ in the underdoped
high temperature superconductors by the Orsay
group~\cite{bobroff,alloulold}. Specifically, they used NMR
measurements to show that the local susceptiblity near a
non-magnetic Li impurity has a $1/T$ divergence as $T \rightarrow
0$. It is expected that in this regime the impurity susceptibility
$\chi_{\rm imp} (T)$, formally defined as the excess contribution
to the uniform magnetic susceptibility due to a single localized
deformation (an ``impurity'') obeys
\begin{equation}
\chi_{\rm imp} (T \rightarrow 0) = \frac{{\cal C}}{k_B T}
\label{chiimp}
\end{equation}
with ${\cal C}= 1/4$. The form (\ref{chiimp}) is very familiar as
the Curie high temperature limit in a large variety of systems.
However, we are interested here in the opposite $T \rightarrow 0$
limit, and then (\ref{chiimp}) can arise only under special
circumstances. (Note also that in any real experiment, there is a
finite density of impurities, and even under the special
circumstances (\ref{chiimp}) will only hold down to some very low
temperature below which interactions between the impurities have
to be accounted for.) We now consider a wide class of correlated
electron ground states and list those that can obey
(\ref{chiimp}) for an arbitrary localized impurity.
\newline
(i) {\em Fermi liquid}: A local spin degree of freedom will obey
(\ref{chiimp}) only if its exchange interaction with the
fermionic quasiparticles is ferromagnetic: then we must have
${\cal C} = S(S+1)/3$ with $S$ integer or half-integer. If the
exchange is antiferromagnetic (as is surely the case in the
cuprates) then Kondo screening removes the moment as $T
\rightarrow 0$ and $\chi_{\rm imp}$ saturates at a finite value.
\newline
(ii) {\em Insulator with N\'{e}el order}: The magnetic order with
$\langle \vec{S} \rangle =0 $ picks a preferred direction in spin
space, and the impurity spin is not free to rotate. The magnetic
susceptibility is finite only in the direction transverse to the
N\'{e}el order, and the pinning of the spin rotation by the
magnetic order leads to a finite $\chi_{\rm imp}$ as $T
\rightarrow 0$. So there is no $S=1/2$ moment in a N\'{e}el
state~\cite{sbv}.
\newline
(iii) {\em Spin gap insulator}: As we have shown in
Fig~\ref{fig4}, it is easy to find configurations (even with
non-magnetic impurities) which have ${\cal C} = S(S+1)/3$ with
$S$ integer/half-integer.
\newline
(iv) {\em Magnetic quantum critical point}: We recently
studied~\cite{sbv} the critical point between (ii) and (iii) in
two dimensions. We found there that (\ref{chiimp}) continues to be
obeyed but ${\cal C}$ is an irrational number---thus, remarkably,
the impurity is associated with an irrational spin.
\newline
(v) {\em $d$-wave superconductor}: This case similar to the Fermi
liquid in that it is possible to have Kondo screening of the
moment as $T \rightarrow 0$. However the conditions for Kondo
screening are much more stringent~\cite{kondo} because of the
lower density of fermionic states at low energy: it is only
present if the antiferromagnetic exchange and the degree of
particle-hole asymmetry exceed some finite value. In the free
moment phase ${\cal C} = S(S+1)/3$ with $S$ integer or
half-integer, while at the impurity quantum critical point between
the free moment and Kondo screened phases ${\cal C}$ is an
irrational number~\cite{mv2}.

The NMR observations show that there is no Kondo screening of the
moment at low doping, and that there is an onset of a non-zero
Kondo temperature ($T_K$) at a critical doping beyond which the
$T_K$ rises rapidly. These results can be very naturally explained
in a model of a $S=1/2$ moment interacting with a $d$-wave
superconductor, as has been shown in recent
work~\cite{tolya,tolya2,mv2}. The separate question of how the
underdoped superconductor has a moment near a non-magnetic
impurity in the first place is answered for ``free'' by the
theoretical framework of Fig~\ref{fig1}. Moving along the
insulating vertical line at $\delta=0$ in Fig~\ref{fig1}, we
proceed from case (ii) to case (iii) via (iv): a precise theory of
the formation of the moment along this route has been
presented~\cite{sbv}. Then we dope the paramagnetic charge-ordered
insulator to reach the superconductor with $\langle \vec{S}
\rangle =0$; along this path the question is not one of the {\em
formation} of the moment, but instead of its {\em quenching} by
the fermionic quasiparticles; the latter issue is precisely that
addressed by the Kondo model. Full translational symmetry must be
restored at some {\em bulk} quantum critical point along this path
(to be discussed in Section~\ref{qc}), but this point will not, in
general, coincide with the {\em impurity} quantum critical point
at which $T_K$ first becomes nonzero.

Next we turn to STM measurements of non-magnetic
impurities~\cite{stm}: these show a sharp low bias peak in the
tunneling conductance at very low $T$ in optimally doped samples.
It was argued in Ref.~\onlinecite{tolya} that this peak is related
to a Kondo resonance in the $d$-wave superconductor and so its
bias and width are determined by $T_K$. An immediate prediction is
that there should be dramatic change in the bias dependence and
spatial distribution of the tunneling conductance as the $T$ is
increased through $T_K$. In particular, the states near the
impurity cross over from a coherent Kondo resonance to ones
determined by potential scattering off the chemically distinct
orbitals on the Zn site.

A more subtle issue, which is far from resolved, is that of the
spatial dependence of the low-bias peak in the low $T$ STM. In
Ref.~\onlinecite{tolya} the author and collaborators proposed that
the unusual spatial dependence may be explained by Kondo
scattering from a spatially diffuse $S=1/2$ moment on the four
nearest neighbor sites of the Zn moment. However, this computation
did not account for the strong potential scattering from the
central Zn site~\cite{bala,zhu}, and also the possible `blocking'
or `filter' effect of the intervening BiO layer between the STM
tip and the CuO$_2$ layer~\cite{zhu2,mbz,oka}. The work of Zhu and
Ting~\cite{zhu}, and our discussions above and in Section~\ref{qc}
instead suggest the following model for future study: over the
time scale of an electron tunneling event it is possible that
local ``stripe'' correlations preferentially pick out a specific
nearest neighbor site of the Zn impurity upon which the moment
resides, just as in Fig~\ref{fig4}. A Kondo model with the moment
localized on this site, along with a strong repulsive potential on
the Zn site, should then be solved for its local density of
states. The actual STM measurements, of course, constitute a long
time average of the tunneling current and this is obtained from
the theory by an {\em incoherent} average of the results of the
four possible positions of the local moment. Note that it is the
relatively long-lived stripe order which demands an incoherent
average, in contrast to the coherent average used in
Ref.~\onlinecite{tolya} which neglected stripe correlations.

\section{Spin and charge ordering quantum transitions}
\label{qc}

We will only consider spin and charge order in an idealized doped
antiferromagnet, where any random disorder potential generated by
the dopant ions is neglected. In the presence of disorder, the
spin order (and concomitant charge order) may appear as spin glass
order~\cite{papa}. Related ideas have been discussed by
Zaanen~\cite{jan}.

\subsection{Charge order}
\label{qc1}

We begin by reviewing simple Landau-theory-like considerations to
characterize charge order in doped antiferromagnets~\cite{zachar}.
We consider charge density wave fluctuations in the $(1,0)$ and
$(0,1)$ directions for a lattice in which the Cu ions occupy the
sites at integer values of $x$, and the O ions occupy half-integer
$x$. Then we can describe the spacetime ($r = (x,y)$) modulations
in the charge density, $\delta \rho (r,\tau)$ by
\begin{equation} \delta \rho (r, \tau) = e^{i K_c x} \phi_x (r,\tau)
 + \mbox{c.c.} + \mbox{$x \rightarrow y$}
\label{rho}
\end{equation}
where $K_c$ is the ordering wavevector and $\phi_{x,y}$ are {\em
complex} order parameters. For commensurate $K_c = 2 \pi m /n$
($m$, $n$, integers) the coupling to the underlying lattice will
allow a term in the action
\begin{equation}
\lambda_c \int d^2 r d\tau \phi_x^n (r, \tau) + \mbox{c.c.} +
\mbox{$x \rightarrow y$}. \label{v}
\end{equation}
This term pins the phase of $\phi_{x,y}$ either at
$\mbox{arg}[\phi_{x,y}] = 0$ for $\lambda_c<0$, or at
$\mbox{arg}[\phi_{x,y}] = \pi/n$, for $\lambda_c>0$ (changes of
$\mbox{arg}[\phi_{x,y}]$ by integer multiples of $2 \pi/n$ are not
significant as they correspond simply to a translation in
(\ref{rho}) by integer lattice spacing); the first case
corresponds to a {\em site-centered} charge density wave, while
the second is {\em bond-centered}. The state in Fig~\ref{fig2}a
has $m=1$, $n=2$ and is bond-centered; note that for this case
only the charge density on all the Cu sites in unmodified---the
modulation is apparent if we examine the charge density on the O
sites, or in off-site correlations of the Cu orbitals which take
the form of a $p_x$ density-wave state~\cite{nayak}.

To choose between the different possibilities for (\ref{rho}) we
have to turn to more microscopic considerations. In the region
with $\langle \vec{S} \rangle =0$ in Fig~\ref{fig1}, a controlled
large $N$ theory~\cite{sr,vojtaprl} was used to follow the
evolution of the $\delta=0$ bond-charge order in Fig~\ref{fig2}a
to non-zero $\delta$: we found only bond-centered waves with $n$
{\em even}. The value of $n$ jumped to a large number at small
$\delta$ and then decreased monotonically through a series of
even integer plateaus with increasing $\delta$ (see the top of
Fig~\ref{fig1}); in particular there was large plateaus at $n=4$
(and occasionally a small plateau at $n=2$) before translational
symmetry was eventually restored to an isotropic $d$-wave
superconductor. (Specific theories of hole doping and fermionic
excitations in states with $n=2$ charge order~\cite{kwon}, and
also with ``plaquette'' order~\cite{assa}, have been presented
recently.) We have proposed that such bond-centered charge order
with even $n$ is characteristic of the high temperature
superconductors in the region with no spin order {\em i.e.}
$\langle \vec{S} \rangle =0 $; in the regime where the ground
state is an isotropic $d$-wave superconductor (see
Fig~\ref{fig1}) strong low energy fluctuations of such bond
charge order can be expected. Some experimental support for this
picture has emerged in recent measurements of the phonon
spectra~\cite{egami,kwon}.

The nature of the quantum theory describing the eventual
restoration of translational invariance has also been
discussed~\cite{vojtaprl}: the critical fixed point between a
$d$-wave superconductor and a $d$-wave superconductor with
co-existing charge order is described by a relativistic field
theory with dynamic exponent $z=1$ which obeys hyperscaling and
exhibits $\omega/T$ scaling. This should be contrasted with the
work of the Rome group~\cite{rome} who have considered charge
ordering transitions in Fermi liquids: such fixed points are very
different and do not satisfy any of the characteristics just
mentioned.

\subsection{Spin order}
\label{qc2}

A related analysis can be developed for spin excitations in the
doped Mott insulator. Here it is convenient to begin in the
paramagnetic Mott insulator at $\delta=0$ which, as shown in
Fig~\ref{fig3}, has a sharp $S=1$ exciton above a spin gap. Upon
moving to the doped superconductor, it is easy to see that this
exciton retains its integrity for a finite range of $\delta$. The
superconductor does have an entirely new class of low energy spin
excitations which are not present in the insulator: these are the
fermionic $S=1/2$ Bogoliubov quasiparticles. However, these low
energy quasiparticles reside only in certain sectors of the
Brillouin zone and momentum conservation constraints can prevent
the $S=1$ exciton from decaying into two $S=1/2$ quasiparticles.
Explicit calculations exhibiting this feature appear in
Ref.~\onlinecite{kwon}. So the main effect of the non-zero
$\delta$ will be renormalize the dispersion of the exciton: in
particular, it is likely that the exciton will eventually move
its minimum away from $(\pi,\pi)$ to a general wavevector
$(\pi,\pi) + K_s$. As in the discussion of the charge density, we
assume that $K_s$ is polarized along $(1,0)$ or $(0,1)$. In the
presence of such excitonic fluctuations, the spin density on the
Cu sites is given by:
\begin{equation}
S_{\alpha} (r, \tau) = (-1)^{x+y} e^{i K_s x} \Phi_{x\alpha}
(r,\tau)
 + \mbox{c.c.} + \mbox{$x \rightarrow y$}
\label{spin}
\end{equation}
where $\alpha=x,y,z$ are the spin directions, and
$\Phi_{x\alpha}$, $\Phi_{y\alpha}$ are complex field operators for
the $S=1$ excitons. We now discuss various terms in the effective
action for $\Phi_{x\alpha}$, $\Phi_{y\alpha}$. For the reasons
just discussed, we will neglect their couplings to the fermionic
quasiparticles, but it is not difficult to include such a coupling
if momentum conservation happens to permit it~\cite{vojtaprl,bfn}.
First, along the lines of Refs.~\onlinecite{vojtaprl,sbv,book} we
have the usual gradient and time derivative terms present in any
quantum antiferromagnet:
\begin{eqnarray}
&& {\cal S}_{\Phi} = \int d^2 r d \tau \Biggl[ |\partial_{\tau}
\Phi_{x\alpha} |^2 + c_1^2 |\partial_x \Phi_{x\alpha}|^2 + c_2^2
|\partial_y
\Phi_{x\alpha} |^2 \nonumber \\
&& + |\partial_{\tau} \Phi_{y\alpha} |^2 + c_1^2 |\partial_y
\Phi_{y\alpha}|^2 + c_2^2 |\partial_x \Phi_{y\alpha}|^2 + s (
|\Phi_{x\alpha}|^2 + |\Phi_{y\alpha}|^2) \nonumber \\ && +
\frac{u_1}{2} \bigl\{ (|\Phi_{x\alpha}|^2)^2 +
(|\Phi_{y\alpha}|^2)^2 \bigr\} + \frac{u_2}{2} \bigl\{
|\Phi_{x\alpha}^2|^2
+ |\Phi_{y\alpha}^2|^2 \bigr\} \nonumber \\
&&~~~+v_1 |\Phi_{x\alpha}|^2 |\Phi_{y\beta}|^2 + v_2
|\Phi_{x\alpha} \Phi_{y \alpha} |^2 + v_3 |\Phi_{x\alpha}^{\ast}
\Phi_{y \alpha} |^2 \Biggr]. \label{sp}
\end{eqnarray}
Second, the underlying lattice may pin $K_s$ to the commensurate
values $K_s=0$ or $K_s = 2\pi/(2p)$ with $p$ integer. For the
special case $K_s=0$, we see from (\ref{spin}) that $\Phi_{x,y}$
have the same spin modulation and so only one of them need be
considered; further they should also be real, and so ${\cal
S}_{\Phi}$ reduces to the familiar O(3) $\varphi^4$ field theory
in 2+1 dimensions~\cite{book}. For $K_s = \pi/p$, as in (\ref{v}),
we have a term in the action
\begin{equation}
{\cal S}_L = \lambda_s \int d^2 r d\tau (\Phi_{x\alpha}^2)^p +
\mbox{c.c.} + \mbox{$x \rightarrow y$} \label{v1}
\end{equation}
which again pins the phase of $\Phi_{x\alpha}$ and $\Phi_{y
\alpha}$ to a discrete set of allowed values. Finally, as noted by
Zachar {\em et al.}\cite{zachar} there is also a coupling between
the spin and charge degrees of freedom
\begin{equation}
{\cal S}_{\Phi\phi} = w \int d^2 r d \tau \phi_x^{\ast}
\Phi_{x\alpha}^2 + \mbox{c.c.} + \mbox{$x \rightarrow y$}
\label{v2}
\end{equation}
which is operative only if $K_c = 2K_s$ and then pins the
long-range spin and charge order to each other. The ordering
observed in LNSCO is of this type with $m=1$, $n=4$, and $p=4$.
Spin correlations pinned at $p=4$ are also universally observed
over a wide doping range in the cuprates: this pinning is
difficult to understand from the semiclassical theory of stripe
formation\cite{zaanen}, and we propose that it is linked via
(\ref{v2}) to singlet bond charge order correlations for which a
large plateau at $n=4$ was found in Ref.~\onlinecite{vojtaprl}.

The above framework can now be used to study the transition into
the magnetically ordered region with $\langle \vec{S} \rangle \neq
0$ in Fig~\ref{fig1}. This transition is driven by decreasing the
value of $s$ in (\ref{sp}) and corresponds to a condensation of
the $S=1$ exciton. The nature of this transition depends crucially
on the role of the charge order, and two distinct cases can be
separated.
\newline
(I) In the first case, long-range charge order remains an innocent
spectator to the spin ordering transition. If the spin ordering is
at a wavevector $K_s = K_c /2$, then the charge ordering, via the
coupling (\ref{v2}), will select $\Phi_{x \alpha}$ over $\Phi_{y
\alpha}$ (say), and also pin its phase so that $\Phi_{x \alpha}$
is real. The theory reduces to that of a single real 3-component
vector, and is again the familiar O(3) $\varphi^4$ field
theory~\cite{book}. This situation is the case along the BC
portion of the spin-ordering phase boundary in Fig~\ref{fig1}.
\newline
(II) In the second case, the spin-ordering fluctuations are
paramount, and charge order appears only in the region with
$\langle \vec{S} \rangle \neq 0$ (assuming the spin correlations
are not spiral); this is the case along the CD portion of the
phase boundary in Fig~\ref{fig1}. The critical theory is now
given by ${\cal S}_{\Phi}+{\cal S}_L$ and the spectator charge
order is determined simply by (\ref{v2}) as $\phi_{x} \sim
\Phi_{x \alpha}^2$, and similarly for $\phi_y$.

Finally, we note the nature of the dynamic spin fluctuation
spectrum near the spin-ordering transition in the region with
$\langle \vec{S} \rangle = 0$. Here, an adequate picture emerges
if we simply treat (\ref{sp}) in the Gaussian approximation, with
$s>0$. A combination of (\ref{sp}) and (\ref{spin}) then implies
the dynamic spin susceptiblity~\cite{bfn}:
\begin{eqnarray}
&& \chi (k, \omega) \propto \frac{1}{c_1^2 (k_x - G_{+x}^x )^2 +
c_2^2 (k_y - G_{+y}^x)^2 + s - (\omega+ i \eta)^2} \nonumber \\
&& ~~~+ \frac{1}{c_1^2 (k_x - G_{-x}^x )^2 + c_2^2 (k_y -
G_{-y}^x)^2
+ s - (\omega+ i \eta)^2} \nonumber \\
&& ~~~+ \frac{1}{c_1^2 (k_y - G_{+y}^y )^2 + c_2^2 (k_x -
G_{+x}^y)^2
+ s - (\omega+ i \eta)^2} \nonumber \\
&& ~~~+ \frac{1}{c_1^2 (k_y - G_{-y}^y )^2 + c_2^2 (k_x -
G_{-x}^y)^2 + s - (\omega+ i \eta)^2} \label{bala}
\end{eqnarray}
where $G_{\pm}^x = (\pi\pm K_s,\pi)$ and $G_{\pm}^y = (\pi,\pi\pm
K_s)$, and $\eta$ is a positive infinitesimal. A related spectrum
has been discussed recently by Batista {\em et al.}~\cite{ortiz}
but on the ordered side of the transition: they have argued that
the superposition of the terms in (\ref{bala}) leads to a maximum
in the spectral density at the commensurate point $k=(\pi,\pi)$
and this is responsible for the ``resonance peak'' in neutron
scattering.

\section{Effect of an applied magnetic field}

In this final section we briefly review recent work~\cite{dsz} on
the effect of an applied magnetic field, oriented perpendicular
to the CuO$_2$ layers, on the above spin and charge ordering
transitions within a superconducting ground state.

So far, we have found that there was little material difference
between the spin/charge ordering transitions of a superconductor
and those of an insulator: the superconductor has additional low
energy quasiparticle excitations, but these often decouple from
the critical spin/charge degrees of freedom because of constraints
from momentum conservation. The situation changes dramatically in
the presence of an external field. Briefly stated, the
``background'' superconducting order has an infinite diamagnetic
susceptiblity and this feeds into an anomalously strong response
of the spin/charge order, much stronger than that would be found
for the corresponding quantum transitions in an insulator. The
effect is quite simple: in the presence of an applied field, $H$,
the type II superconductor reacts by setting up superflow currents
which screen the field. A standard calculation using the standard
Ginzburg Landau free energy shows that
\begin{equation}
\mbox{mean kinetic energy density of superflow} \sim
\left|\frac{H}{H_{c2}}\right| \ln \left|\frac{H_{c2}}{H} \right|.
\label{ke}
\end{equation}
This shift leads to concomitant change in the effective action for
the spin/charge order~\cite{dsz} via the simple Landau theory
coupling (emphasized in Ref.~\onlinecite{zhang})
\begin{equation}
{\cal S}_{sc} = \kappa \int d^2 r d \tau |\psi|^2
|\Phi_{x\alpha}|^2 + \mbox{$x \rightarrow y$}
\end{equation}
(and similarly for $\phi_{x,y}$) to the superconducting order
parameter $\psi$. Crudely speaking, this correction is accounted
for an $H$-dependent shift in the value of $s$ in (\ref{sp}) which
has the form~\cite{dsz}
\begin{equation}
s(H) = s - \tilde{\kappa} \left|\frac{H}{H_{c2}}\right| \ln
\left|\frac{H_{c2}}{H} \right|. \label{sh}
\end{equation}
This simple expression now has a number of strong consequences for
the position of the phase boundary for the onset of long-range
spin order, and for the nature of the spin excitation spectrum on
either side of the phase boundary. On the side with long-range
spin order, (\ref{sh}) implies that the Bragg elastic scattering
intensity should increase (for $\widetilde{\kappa} > 0$)
as~\cite{dsz} $\sim (|H|/H_{c2}) \ln (H_{c2}/|H|)$: this
prediction is consistent with recent experiments~\cite{younglee}.
On the side with no spin order, the dynamic susceptibility will be
given by (\ref{bala}) but with $s$ replaced by $s(H)$: hence the
minimum excitonic energy, $\sqrt{s(H)}$, will decrease with
increasing $H$---this trend is also consistent with
experiments~\cite{lake}.

All of the above effects considered the uniform, spatially
averaged, consequence of the applied field. However, as is well
known, the superflow kinetic energy in (\ref{ke}) is
inhomogenously distributed in the form of a vortex lattice, and
this has interesting observable consequences for the spin/charge
order. In the spin-ordered phase, there are weak satellite Bragg
peaks~\cite{arovas,dsz} around $G_{\pm}^{x,y}$, separated from
$G_{\pm}^{x,y}$ by reciprocal lattice vectors of the vortex
lattice. In the region without spin order ($\langle \vec{S}
\rangle = 0$) which was considered in Ref.~\onlinecite{dsz}, the
wavefunction, $\Phi_{x,y\alpha} (r)$, of the lowest energy
exciton (which is an excited state at energy $\sqrt{s(H)}$) will
be peaked at the vortex core, but will extend will outside the
vortex core to a distance $\sim c_{1,2}/\sqrt{s(H)}$ (up to the
natural maximum of the vortex lattice spacing), where the
velocities $c_{1,2}$ are expected to be of order the spin-wave
velocity in the antiferromagnetic state at $\delta=0$. Given the
large size of the exciton wavefunction, small local disorder may
well enhance $\Phi_{x \alpha}$ over $\Phi_{y \alpha}$ (or vice
versa) in a given set of vortices. Associated with this excitonic
excited state is a spin modulation specified by $\Phi_{x,y\alpha}
(r) $ and (\ref{spin}), and a charge modulation given by
$\phi_{x,y}(r) \sim \Phi_{x,y\alpha}^2 (r)$ and (\ref{rho}).

\acknowledgements The ideas reviewed here were developed in work
with Eugene Demler, Kwon Park, Anatoli Polkovnikov, Matthias
Vojta, and Ying Zhang, and I thank them for fruitful
collaborations. I thank M.~Vojta, J.~C.~Davis, and S.-C.~Zhang for
helpful comments on the manuscript. This research was supported by
US NSF Grant DMR 0098226.



\end{document}